# Dynamically Weighted Ensemble-based Prediction System for Adaptively Modeling Driver Reaction Time

Chun-Hsiang Chuang, *Member IEEE*, Zehong Cao, *Member IEEE*, Po-Tsang Chen, Chih-Sheng Huang, Nikhil R. Pal, *Fellow IEEE*, Chin-Teng Lin, *Fellow IEEE*

*Abstract*—Motor vehicle crashes are the leading cause of fatalities worldwide. Most of these accidents are caused by human mistakes and behavioral lapses, especially when the driver is drowsy, fatigued, or inattentive. Clearly, predicting a driver's cognitive state, or more specifically, modeling a driver's reaction time (RT) in response to the appearance of a potential hazard warrants urgent research. In the last two decades, the electric field that is generated by the activities in the brain, monitored by an electroencephalogram (EEG), has been proven to be a robust physiological indicator of human behavior. However, mapping the human brain can be extremely challenging, especially owing to the variability in human beings over time, both within and among individuals. Factors such as fatigue, inattention and stress can induce homeostatic changes in the brain, which affect the observed relationship between brain dynamics and behavioral performance, and thus make the existing systems for predicting RT difficult to generalize. To solve this problem, an ensemble-based weighted prediction system is presented herein. This system comprises a set of prediction sub-models that are individually trained using groups of data with similar EEG-RT relationships. To obtain a final prediction, the prediction outcomes of the sub-models are then multiplied by weights that are derived from the EEG alpha coherences of 10 channels plus theta band powers of 30 channels, whose changes were found to be indicators of variations in the EEG-RT relationship. The results thus obtained reveal that the proposed system with a time-varying adaptive weighting mechanism significantly outperforms the conventional system in modeling a driver's RT. The adaptive design of the proposed system demonstrates its feasibility in coping with the variability in the brain-behavior relationship. In this contribution surprisingly simple EEG-based adaptive methods are used in combination with an ensemble scheme to significantly increase system performance.

## I. INTRODUCTION

D RIVING safety always attracts public attention. According to

C. H. Chuang is with Department of Computer Science and Engineering, National Taiwan Ocean University, Taiwan, and Faculty of Engineering and Information Technology, University of Technology Sydney, Australia (e-mail: chchuang@ieee.org)

Z. Cao is with Discipline of ICT, School of Technology, Environments and Design, University of Tasmania, TAS, Australia (email: Zehong.Cao@utas.edu.au)

P.T. Chen and C.S. Huang are with the Brain Research Center, National Chiao Tung University, Taiwan (e-mail: hendry5628@gmail.com; chih.sheng.huang821@gmail.com)

N. R. Pal is with the Institute of Electrical Control Engineering, Indian Statistical Institute, India (e-mail: nrpal59@gmail.com)

C.T. Lin is with Faculty of Engineering and Information Technology, University of Technology Sydney, NSW, Australia (email: Chin-Teng.Lin@uts.edu.au)

World Health Organization, about 1.25 million people are reported to be killed annually and between 20 and 50 million people are non-fatally injured as a result of road traffic accidents [1]. Drivers' behavioral errors and lapses are responsible for 90% of road accidents [2]. Despite advances in the development of advanced driver assistance systems, accident countermeasures have not kept pace with the growing number of accidents on the road. Unless urgent action is taken to improve the safety of driving, road traffic injuries may become the seventh leading cause of death across all age groups [3].

Drowsy or fatigued driving is a dangerous behavior that causes many fatal and non-fatal accidents every year [4]. Strictly speaking the exact biological processes behind drowsiness and fatigue may be different but as far as our problem is concerned, their impact on driving performance is the same as fatigue leads to drowsiness. We do not make an explicit distinction between the two. However, since we use several Gaussian Mixture Models (GMMs), if the EEG signatures of the two are different, they would be implicitly taken care by some components of the GMMs. This is a unique advantage of our proposed framework. Drivers with impaired driving ability are less likely to maintain a stable lane position, and their vehicles tend to drift or swerve. If a small deviation from the lane center is not corrected rapidly using the steering wheel, then the vehicle may soon collide with other vehicles in the adjacent lanes or run off the road in just a few seconds. Drowsiness and fatigue reduce the driver's attention and the response time to traffic events and emergencies. Thus, modelling driver's reaction time (RT) to traffic emergencies can help to monitor drowsiness.

Previous studies [5-8] have shown a strong connection between drowsiness and behavioral lapses before an accident. Mainstream techniques for detecting drowsiness are broadly divided into two major categories. Some focus on the behavior of drivers such as movements of the steering wheel or the deviations of the moving vehicle from the center of a lane [9]. Others focus on physiological features, such as brain waves [10, 11], eye movements, respiratory rate and heart rates [12, 13]. Over the last two decades, the electric field that is produced by the activities in the brain, recorded by an electroencephalogram (EEG), has been proven to be a robust physiological indicator of human cognitive states. EEG is the only brain imaging modality with a high temporal and fine spatial resolution that is sufficiently lightweight to be worn in operational settings [14]. Many studies have demonstrated the feasibility of accurately estimating shifts in a driver's level of alertness from changes in his or her performance in various tasks, from a simple auditory target detection task to a complex driving task, by monitoring changes in the EEG power spectrum [15]. Changes in the EEG theta (4-7 Hz) and alpha (8-12 Hz) activity are strongly correlated with changes in cognitive performance [16, 17]. Most relevant studies have reported significant increases in the theta power, the frequency of theta burst or the duration of episodes of theta activity on the EEG when the cognitive state changes from alert to poor/drowsy, during prolonged driving, or a progressive deterioration of the driver's vigilance [18-20]. However, results concerning alpha activity vary across studies. Analyzing the ratio of theta power to alpha power [21, 22] suggests that alpha activity vanishes and is replaced by theta activity during micro sleep episodes. In contrast, the alpha power, the ratio of theta+alpha and beta power, or the ratio of alpha and beta power increase as driving error increases and fatigue [23-25] occurs. Some studies [26] have found that power in the alpha band varies (basically follows an inverted-U shaped curve) as behavioral performance worsens. These differences in results may arise from large variations across these studies in tasks, experimental design and setup, behavioral measures, baseline brain activity and other factors, making generalization difficult.

Many EEG-based driver assistance systems [12, 13, 20, 27-34] have been proposed for lapse detection, fatigue monitoring, alertness evaluation, mind-wandering, or accident prevention, based on characterizing the relationship between EEG features and cognitive state or behavioral alertness, such as a driver's RT to the appearance of a potential hazard. These systems use statistical models or machine learning approaches, such as Gaussian mixture model [35], the support vector machines (SVMs) [27], and neural networks [30] as the core algorithm. For example, the method for automatically detecting drowsiness based on spectral and wavelet analysis has a classification accuracy of 87.4% and 83.6% for alertness and drowsiness, respectively [36]. The SVM-based models predict the transition from alertness to drowsiness with a reliability of approximately 90% [37, 38]. The studies [39-41] conducted so far show that constructing neural fuzzy inference network systems and support vector regression (SVR) systems with a radial basis function (RBF) kernel using EEG power spectra and independent component analysis achieve a satisfactory accuracy and predictive performance in both vigilance detection and behavioral lapse detection. These demonstrate the feasibility of using EEG with an advanced algorithm in modelling the brain-behavior relationship and detecting fatigue, drowsiness, or behavior lapses during driving.

However, most of the existing systems are based on a single optimal prediction model that is applied to all circumstances. The evidence reveals that neurological signals and behavior performance vary amongst individuals and over time, challenging the development of a fixed model [42]. Thus, the results derived from these systems may not be accurate due to wide variation of brain behavior of the general population.

These systems also lack robustness as they do not consider the variability of human beings and hence result in the inaccurate monitoring of driver's reaction time and human cognitive states. To address the stated concerns, this work proposes an adaptive design for a monitoring system. The proposed system integrates an ensemble scheme with a time-varying adaptive weighting mechanism and significantly outperforms the conventional systems in modeling driver's RT. The ability of this system to handle versatility in the brain-behavior model is demonstrated.

## II. MATERIALS AND METHODS

### A. Participants

In this study, 33 subjects (including 2 females) aged 20–30 years with normal or corrected-to-normal vision were recruited to participate in a sustained-attention driving experiment. The subjects were required not to have imbibed alcoholic or caffeinated drinks or to have participated in strenuous exercise the day before the experiments were performed to ensure that their driving performance could be accurately evaluated. All subjects were informed about the experimental procedure and the driving task. Before the experiments, all subjects practiced driving in the simulator to become acquainted with it and with the experimental procedure. All subjects were asked to read and sign an informed consent form before participating in the EEG experiments. The study was approved by the Institutional Review Board of the Veterans General Hospital, Taipei, Taiwan.

### B. Experimental Environment and Paradigm

This study implemented an immersive driving environment (Fig. 1A) [43-45], which provided an immersive experience of simulated nighttime driving on a four-lane highway at a fixed speed of ~100 km/h. This high-fidelity interactive highway scene was developed by WorldToolKit (WTK) engine. The driving environment consisted of a driving simulator and six scenes, which were supported by six identical projectors and PCs running the same driving program. One of the PCs, which recorded the driving parameters like steering wheel angle and vehicle position, was connected with all the other PCs by local area network (LAN) to keep all the streams in a well synchronized manner. All the parameters were automatically registered in a log file and synchronized with EEG data recorder (see Section II-C).

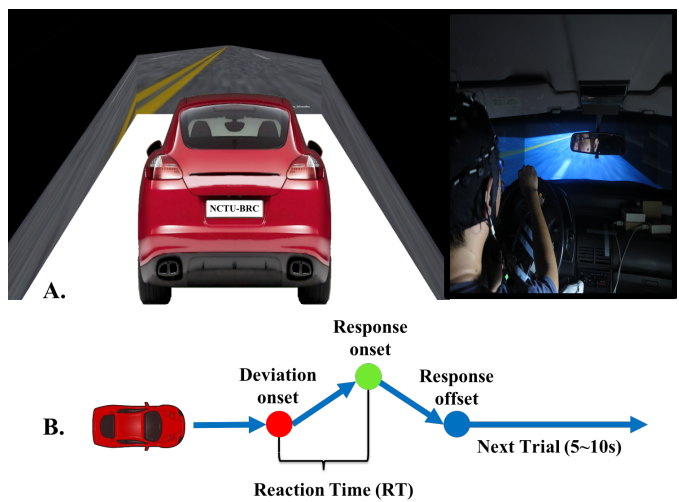

Fig. 1. Simulated driving task. (A) Immersive driving experiment environment. (B) Event-related lane-departure paradigm.

Regarding the experimental paradigm, lane-departure events were randomly activated during the simulated driving to cause the car to drift away from the center of the cruising lane (deviation onset). The subjects were instructed to steer the car back (response onset) to the lane center (response offset) as soon as possible after becoming aware of the deviation. The design of this lane-keeping task is inspired by psychomotor vigilance task which is a well-known sustained-attention task used to measure the speed with which subjects respond to a visual stimulus [46]. In our experiment, the lane drifts were randomly introduced by the scenario computer. Driver's reaction times in response to these events were calculated to assess their alertness. When a subject is in a drowsy state, he/she is unlikely to respond to the lane-departure event. In that case, the vehicle would hit the left and right curb of the roadside within 2.5s and 1.5s, respectively. Until a corrective action is taken by the subject, the vehicle would continue to move along the curb. The inter-trial interval was set to 5-10 s.

The lapse in time between the onset of deviation and response was defined as the reaction time (RT) (Fig. 1B). The subject's cognitive state and task engagement were monitored via a surveillance video and the vehicle trajectory to decipher the attentiveness of the subject, which was used to reject the trials when the steering wheel was not synchronistic with the visual effect. The driving experiment was conducted in the early afternoon (13:00–15:00) after lunch because the circadian rhythm of sleepiness was at its peak at noon [47, 48]. Additionally, the highway scene was monotonous and the task demand was low and hence were likely to induce drowsiness [43, 49]. Under such conditions, subjects had difficulty regulating attention and performance [50], resulting in long RTs. This experiment enabled a wide range of RTs to be collected in response to traffic events. The experiment lasted for 60-90 minutes. The duration of the driving experiment varied because the experimenters sometimes prolonged an experiment time to collect long RTs. The average number of lane-departure events per experimental run was approximately 400. A total of 73 experimental sessions were collected from 33 subjects who participated in the experiment multiple times (1-6 times).

### C. Data Collection and Preprocessing

During the experiments, the EEG signals were recorded using Ag/AgCl electrodes that were attached to a 32-channel Quik-Cap (Compumedical NeuroScan). Thirty electrodes were arranged according to an extended international 10-20 system, i.e., in addition to the standard 21 electrode positions, another extra 11 positions were used. These 32 electrodes included two reference electrodes which were placed on both mastoid bones. The skin under the reference electrodes was abraded using Nuprep (Weaver and Company, USA) and disinfected with a 70% isopropyl alcohol swab before calibration. The impedance of the electrodes was calibrated under 5kΩ using NaCl-based conductive gel (Quik-Gel, Neuromedical Supplies®). The EEG signals from the electro-cap were amplified using the Scan NuAmps Express system (Compumedics Ltd., Australia) and recorded at a sampling rate of 500 Hz with a 16-bit quantization [51]. Before the data were analyzed, the raw EEG recordings were pre-processed using a digital band-pass filter (1-30 Hz) to remove line noise and artifacts.

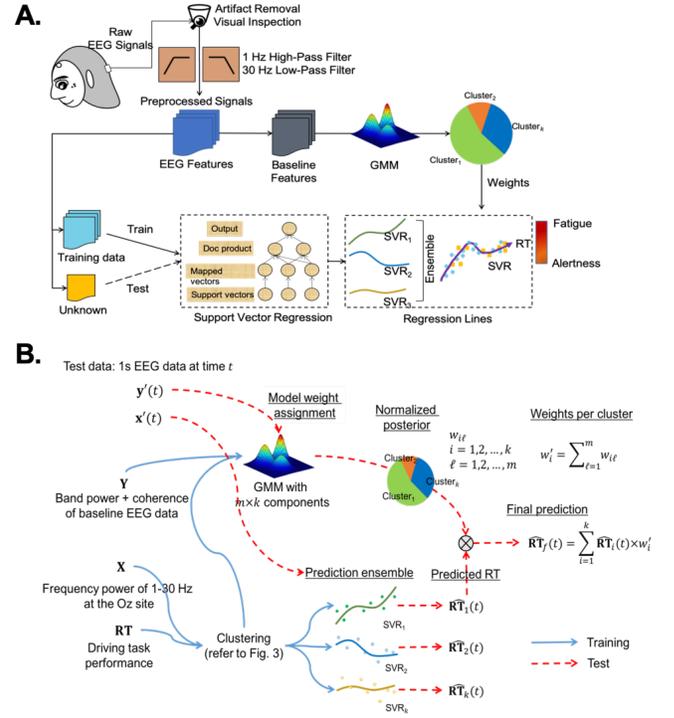

Fig. 2. (A) Acquisition of EEG signal, its processing, and analysis along with use of an ensemble of regression models (B) The framework of a dynamically weighted ensemble-based prediction system for adaptive prediction of driver's reaction time (RT), where the blue and red arrows represent the training and test processes, respectively. Multiple models were used to characterize EEG-RT relationships. The band powers and coherences extracted from EEG data were used to train Gaussian mixture model (GMM) for assigning the weight of the corresponding model. EEG powers of 1-30 Hz at the Oz channel were used to train support vector regression (SVR) for predicting RTs. RTs estimated using various SVRs are then multiplied by weights estimated using GMM to yield final prediction.

## III. ENSEMBLE-BASED PREDICTION SYSTEM FOR ADAPTIVELY MODELING DRIVER REACTION TIME

A Gaussian mixture model assumes that data points are generated from a mixture of a finite number of Gaussian distributions with unknown parameters. The mixture models can be considered a generalization of k-means clustering which can incorporate the covariance structure of the data (in this case of the EEG powers) along with the means of the Gaussians. Furthermore, the slow behavioral response of lane lapses induced by fatigue correlates with the changes of EEG activities. To link EEG powers with RTs, a nonlinear model can incorporate both the linear and nonlinear relationships between EEG power spectra and RTs. The SVM is a popular approach to model the multidimensional function estimation problem, such as regression and classification. When used to solve the function approximation and regression estimation problems, SVM is denoted as the SVR. Figure 2-A shows the system for acquisition of EEG signal, its processing, and analysis, and use of an ensemble of regression models, where EEG and behavior are modelled using SVR, and the predicted outputs are converted to different levels of fatigue.

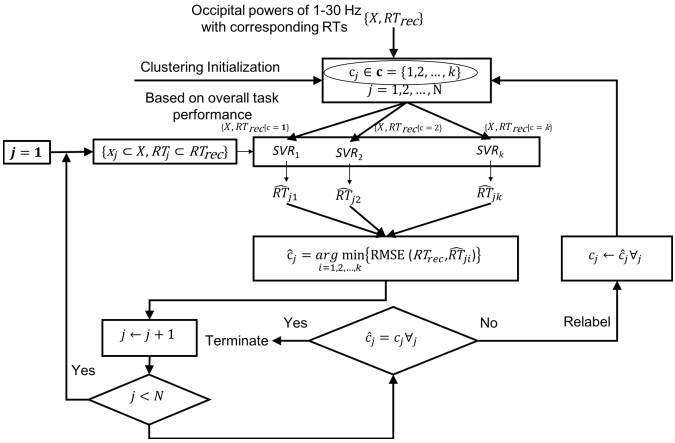

Fig. 3. Recursive procedure for clustering data with a similar EEG-RT relationship. Given $k$ clusters, the data of the $j$th subjects including EEG data and the corresponding RTs were initially labeled as $c_j$ based on driver's overall task performance. Then, the data within the same cluster were used to train a single SVR. The label of data was changed if the minimum RMSE was achieved by the SVR trained using the data of a different group. This recursive mechanism terminates when labels of all data were fixed.

Given that $p(\mathbf{x}, \mathbf{RT})$ specifies the relationship between the RTs of drivers and their accompanying patterns of brain activity $\mathbf{x}$, most previous works have sought a specific prediction model $g: \mathbf{x} \rightarrow \widehat{\mathbf{RT}}$ that fits this relationship. To equip a prediction model with the capability of adaptively modeling a driver's RT in response to a traffic event, this work proposes a new prediction system, as presented in Fig. 2-B, integrating an ensemble scheme with a time-varying adaptive weighting mechanism. To this end, such a mapping is assumed to be specified by a set of prediction models $\mathcal{G}$, where the number of models is set to $k$. Specifically, $k$ prediction models (Model$_1$, Model$_2$, ..., Model$_k$) were constructed using EEG spectral powers of 1-30 Hz at the Oz channel and their corresponding RT. We use the brain activity of the Oz channel to train the prediction model because a review of the literature [52] gives a strong evidence that the brain activity at the occipital region is highly associated with human alertness/drowsiness during driving. With respect to the mechanism for assigning model weights, the EEG band powers accompanying with coherences were used to train k Gaussian mixture models (GMMs). Each GMM corresponds to one prediction model and the GMM provides a dynamic weight for its associated prediction model that is used in the ensemble aggregation. The posterior probability of every 1s EEG data at time $t$ was normalized and considered weights of the models. Finally, the prediction output was obtained by a weighted averaging.

## A. Data Processing

All frequency responses were captured using a window of 90 secs (45000 points). The EEG signal was estimated using 512-point fast Fourier Transformation (FFT) following a doubling of the length of the time signal by zero padding. The step size was set to 1 sec (500 points). Specifically, each 90-sec sliding window of the EEG signal was segmented in several 512-points sub-windows. The sub-windows (512 points) were expanded to 1024 points (using the zero-padding) method to increase the resolution of the power spectra. Each 512-points sub-window were then transformed to the frequency domain using 512-points FFT. The mean value of all sub-windows in the frequency domain was calculated as the output of the FFT process. Finally, the frequency responses were converted into a dB scale, yielding a logarithmic spectral power with 120 frequency bins from ~1 to ~30 Hz. The spectral powers of 1-30 Hz in the Oz channel were used to construct the prediction ensemble. The theta band (4-7 Hz) powers of all channels and the coherence between pairs of channels, which was calculated by phase locking value [53], were used to construct the GMM.

## B. Model Clustering

The first step in implementing the proposed adaptive model was to obtain the model cluster. An iterative procedure, presented in Fig. 3, was proposed to cluster data with similar EEG-RT relationships. Given Subject $j$, his or her EEG data $\mathbf{x}_j$ and the corresponding task performance data $\mathbf{RT}_j$ collected during the experiment were assigned an initial label $c_j \in \mathbf{c} = \{1, 2, ..., k\}$, where $j = 1, 2, ..., N$ and $N$ is the total number of subjects. The data belonging to the same cluster, i.e., $\{\mathbf{X}, \mathbf{RT_{rec}}|c = i\}$, where $i = 1, 2, ..., k$, were used to train a single SVR. Then, the data was relabeled as $\hat{c}_j = \underset{i=1,2,...,k}{arg\ min}\{\text{RMSE}\,(\mathbf{RT_{rec}}, \widehat{\mathbf{RT}}_{ji})\} \in \hat{\mathbf{c}}$, where the RMSE represents the root mean squared error. Then we repeat the process of training SVRs and reassignment of labels. This iterative procedure is terminated when no further changes take place in the labeling, indicating that the data within the same cluster tended to have a similar EEG-RT relationship.

By observing the recorded RT of all data, as presented in Fig. 4, the overall task performance of subjects could be divided into three major clusters (optimal, suboptimal, and poor), that is $k = 3$ in this study. The task performance of each session was defined as the ratio of the recorded RT to the mean of the fastest 10% of RTs. Thirty-six sessions performed mostly at an optimal level (i.e., RT ratio $\leq 2$) throughout the experiment, while 14 and 23 sessions tended to exhibit suboptimal ($2 <$ RT ratio $\leq 3$) and poor (RT ratio $> 3$) task performance, respectively.

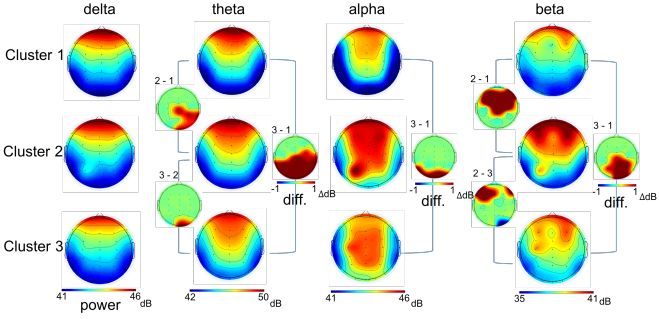

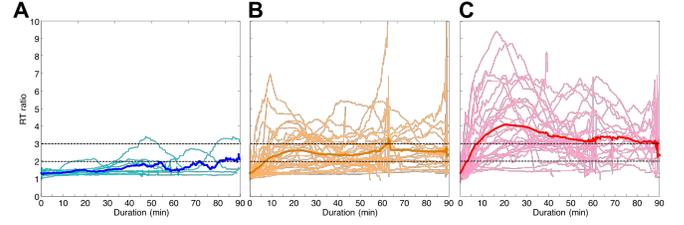

Fig. 4. Temporal changes of RTs in response to lane departure events. (A) The group optimal task performance. (B) The group with suboptimal task performance. (C) The group with poor task performance.

Fig. 5. Topographic maps of the EEG delta, theta, alpha, and beta powers. Cluster 1 - Group with optimal task performance. Cluster 2 - Group with suboptimal task performance. Cluster 3 - Group with poor task performance. The small subplots between scalp maps indicate the significant differences between two clusters.

### C. Dynamically Weighted Ensemble

As presented in Fig. 2-B, an SVR-based prediction ensemble was integrated with a GMM-based model weight assignment to adaptively model a driver's RT. Given $k$ clusters of data extracted from GMM, the proposed weight assignment mechanism used the EEG band power of all channels and coherence of ten selected channels (O1, O2, Oz, P3, Pz, P4 ,CP3 ,CPz ,CP4, and Cz) at time $t$ to evaluate the model weights every second. This weight assignment can be represented as a mapping $f : \mathbf{Y} \to w_{i|i=1,2,\dots,k} \in [0,1]$ , where the outcome (posterior probability) $w_i$ is a degree of belonging to $c_i$. In this study, GMM was applied to the first five minutes of the data to construct this map, where $m$ components were constructed for each cluster. In the prediction process, the normalized posterior probability of the data $\mathbf{y}'(t)$ at time $t$, $w_{i\ell}(t)$, was calculated every second and then summed up across $m$ components, $w_i' = \sum_{\ell=1}^{m} w_{i\ell}(t)$, so as to obtain the weights of the $i^{\text{th}}$ models, where $\ell = 1, 2, \dots, m$.

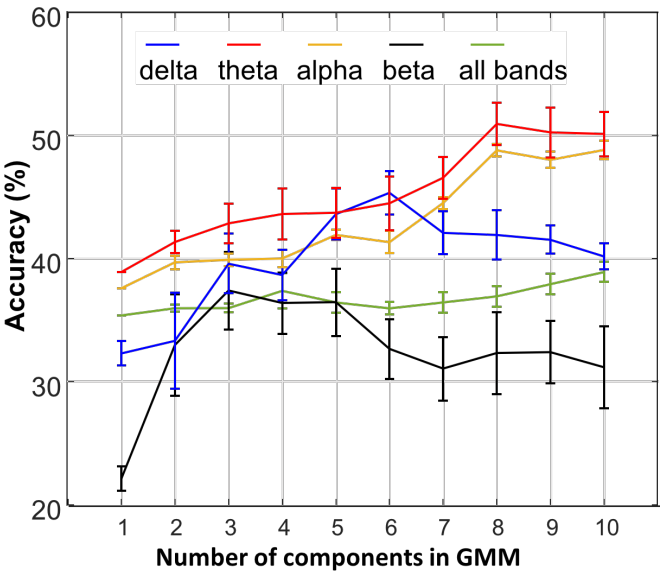

Fig. 6. Classification accuracies obtained using $m$ Gaussian components for each cluster and various frequency bands.

Regarding the prediction ensemble, $k$ SVRs were individually trained using the dataset of group $c_i$ , i.e., $\{\mathbf{X}, \mathbf{RT}_{\mathrm{rec}}|c = i\}$ , where $i = 1, 2, \dots, k$ , where $\mathbf{X}$ is the EEG frequency power of 1-30 Hz at the Oz channel and $\mathbf{RT}_{\mathrm{rec}}$ denotes the recorded RT. During the prediction, the RT of each EEG data at time $t$, $\mathbf{x}'(t)$, was evaluated by submitting it to each SVR of the ensemble to obtain the predictive outcomes $\widehat{\mathbf{RT}}_i(t)$. Then, the final prediction was obtained by $\widehat{\mathbf{RT}}_f(t) = \sum_{i=1}^{k} \widehat{\mathbf{RT}}_i(t) \times w_i'$. In this work, SVR with an RBF [54] kernel was adopted and $k = 3$.

## IV. RESULTS

### A. Differences of EEG Spectral Dynamics Among Groups

The experiment was conducted using three SVR models with different weights to characterize variations in the EEG-RT relationship. The weights were calculated using GMM. Fig. 5 presents the topographic maps of EEG delta, theta, alpha, and beta powers associated with three clusters, where clusters 1, 2, and 3 represent the groups with optimal, suboptimal, and poor task performance, respectively. The obtained results revealed that the EEG powers of theta and beta bands significantly varied among clusters (Wilcoxon signed-rank test, $p<0.05$). For the theta band, the power of the posterior region, especially for the occipital area, increased significantly as Cluster varied from 1 to 3 (see the small subplots between scalp maps). With respect to the beta power, the power of the anterior region differed significantly between Clusters 1 and 2, as well as between Clusters 2 and 3. Specifically, the power of Cluster 2 was significantly larger than that of Clusters 1 and 3. Compared to Cluster 1, the beta power of Cluster 3 was significantly larger at the posterior region. Some significant differences in the alpha power between Clusters 1 and 3 were also identified (Cluster 3 > Cluster 1). This variation of EEG patterns among clusters suggests that the EEG power spectrum effectively indicates changes in the EEG-RT relationship.

TABLE 1
CLASSIFICATION ACCURACY IN DATA CLUSTERING

|  |  |  | TEST DATA |  |
|---|---|---|---|---|
|  |  | Model$_1$ | Model$_2$ | Model$_3$ |
| $\delta$ ($m = 6$) | Cluster 1 | 68.8±5.9 | 23.3±7.3 | 8.3±1.6 |
|  | Cluster 2 | 68.9±2.1 | 14.4±6.7 | 16.7±20.1 |
|  | Cluster 3 | 64.0±10.4 | 20.9±7.0 | 15.1±4.6 |
| $\theta$ ($m = 8$) | Cluster 1 | **75.8±1.4** | 13.6±2.2 | 10.5±0.8 |
|  | Cluster 2 | 33.7±1.4 | **32.7±2.4** | 33.4±1.4 |
|  | Cluster 3 | 20.5±4.5 | 23.5±2.6 | **55.9±1.0** |
| $\alpha$ ($m = 8$) | Cluster 1 | 64.6±7.1 | 32.2±5.6 | 30.8±1.6 |
|  | Cluster 2 | 58.8±7.0 | 39.8±8.8 | 12.7±6.2 |
|  | Cluster 3 | 30.4±2.2 | 66.0±2.2 | 34.2±2.5 |
| $\beta$ ($m = 3$) | Cluster 1 | 65.5±5.5 | 14.6±5.7 | 19.8±7.3 |
|  | Cluster 2 | 74.4±8.7 | 14.0±9.7 | 11.6±8.2 |
|  | Cluster 3 | 69.1±7.0 | 17.1±6.9 | 13.9±7.3 |
| $\delta, \theta, \alpha, \beta$ ($m = 10$) | Cluster 1 | 50.9±5.0 | 10.1±3.3 | 39.3±4.9 |
|  | Cluster 2 | 53.3±5.2 | 28.0±4.6 | 18.7±4.6 |
|  | Cluster 3 | 45.8±4.5 | 20.8±3.3 | 33.3±2.0 |

(TRAINING DATA)

### B. Accuracy of Model Clustering using EEG Band Power

A multidimensional feature vector consisting of different band power of 30 channels are initially used to construct the GMM. To obtain a suitable value for the number of components $m$ in the GMM and to determine the spectral band that should be used to achieve a satisfactory accuracy, the classification accuracy was calculated across $m = 1,2,...,10$ and various bands by the maximum a posteriori (MAP) estimation and the leave-one-subject-out validation. As shown in Fig. 6, the results reveal that the classification accuracy varied with $m$ and the frequency bands. The best classification accuracies for the delta, theta, alpha and beta bands, as well as for the combination of the four bands were 46±2%, 51±10%, 49±10%, 38±6%, and 39±10%, respectively. These best performances for delta, theta, alpha and beta occurred when $m = 6$, $m = 8$, $m = 8$, $m = 3$, and $m = 10$, respectively. Overall, the performance of the theta band was significantly better than that of others (Wilcoxon signed-rank test, FDR-adjusted $p$-value<0.01).

Table 1 presents the confusion matrices of the classification accuracy of the GMM which is used to model clusters with various frequency bands data. For the delta band ($m = 6$), the classification accuracies of Model$_1$, Model$_2$, and Model$_3$ were 68.8±5.9%, 14.4±6.7% and 15.1±4.6%, respectively, revealing that Model$_2$ and Model$_3$ using the delta power extensively misclassified data. With respect to the theta band ($m = 8$), the classifications accuracies of Model$_1$, Model$_2$, and Model$_3$ were 75.8±1.4%, 32.7±2.4%, and 55.9±1.0%, respectively. These results revealed that Model$_1$ and Model$_3$ using the theta power classified more samples, compared to that using EEG powers in other frequency bands. For the alpha band ($m = 8$), the correct classification rates of Model$_1$, Model$_2$, and Model$_3$ were 64.6±7.1%, 39.8±8.8%, and 34.2±2.5%, respectively. For the beta band ($m = 3$), the classification accuracies of Model$_1$, Model$_2$, and Model$_3$ were 65.5±5.5%, 14.0±9.7%, and 13.9±7.3%, respectively, revealing that like the delta band, here also Model$_2$ and Model$_3$ easily misclassified data. Table 1 also reveals that the use of combination of the four bands ($m = 10$) also does not improve the performance of Model$_2$ and Model$_3$. Taken all together, theta power is more suitable for clustering fatigue levels instead of the delta, alpha, beta or combined EEG powers. The classification results suggest that the theta band power is a useful feature in constructing weight assignment in the GMM.

### C. Accuracy of Model Clustering using EEG Band Power plus EEG Coherence

Figure 7 shows the classification accuracy of the GMM models with EEG theta power and EEG coherence in grouping EEG baseline data. Overall, the EEG features of the theta band power accompanying with the coherence features are found to improve the classification performance. These results reveal that the best performance (68.8%) is achieved by adding the alpha coherence and using 15 components for each cluster in the GMM. Table 2 further demonstrates that the improved classification accuracies of Model$_1$, Model$_2$, and Model$_3$ are 78.7±4.3%, 46.4±7.2% and 69.6±3.2%, respectively.

### D. Prediction of Driver's Reaction Time

In this work, the ensemble system was set to consist three prediction models which were trained using the data from different clusters. Each ensemble member was trained using EEG spectral powers of 1–30 Hz at the Oz channel. The leave-one-subject-out validation method was used to evaluate system performance. That is, the ensemble system was trained using data 72 sessions and was validated using data from the rest of the sessions. To verify the superiority of the proposed model, the root mean square errors (RMSE) between the recorded RT ($\mathbf{RT}_{rec}$) and the RTs that were predicted using a single SVR ($\widehat{\mathbf{RT}}_{sgl}$), an ensemble with a fixed weight ($\widehat{\mathbf{RT}}_{W}$), and an ensemble with a dynamically adaptive weight ($\widehat{RT}_{dynW}$) were calculated and compared. The smaller RMSE, the higher prediction accuracy.

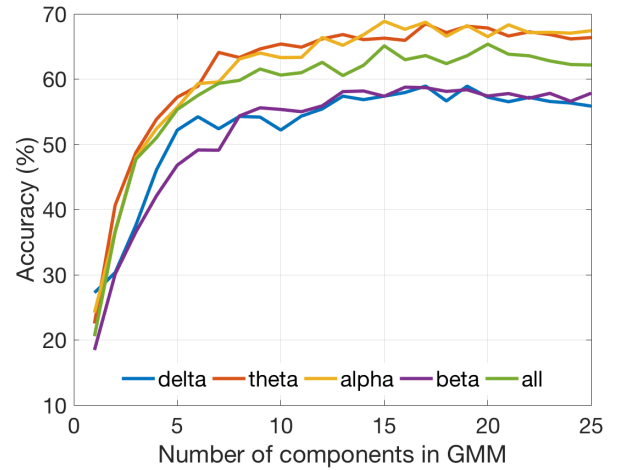

Fig. 7. Classification accuracies of various numbers of Gaussian components obtained using EEG theta powers of 30 channels with EEG coherence of 10 selected channels (O1 , O2 , Oz , P3 , Pz , P4 , CP3 , CPz , CP4 , Cz) in delta, theta, alpha, and beta bands.

TABLE 3
RMSE OF PREDICTION RESULT. THE ASTERISK INDICATES THE SIGNIFICANT DIFFERENCE OF RMSE BETWEEN SINGLE AND THE PROPOSED ENSEMBLE MODELS.

| | MODEL | | |
|---|---|---|---|
| | Single | Ensemble (fixed weights) | Ensemble (dynamic weights) |
| Cluster 1 | 1.19±0.78 | 1.20±0.46 | 1.12±0.53 |
| Cluster 2 | 2.23±0.93 | 1.88±0.72 | 1.36±0.68* |
| Cluster 3 | 2.37±0.81 | 1.79±0.62 | 1.56±0.52* |

TABLE 4
RMSE OF REACTION TIME PREDICTION USING DIFFERENT MODELS

| | | TEST DATA | | |
|---|---|---|---|---|
| | | Cluster 1 | Cluster 2 | Cluster 3 |
| TRAINING DATA | Cluster 1 | **0.93±0.43** | 1.74±0.49 | 1.71±0.45 |
| | Cluster 2 | 2.26±0.75 | **1.33±0.32** | 2.34±0.46 |
| | Cluster 3 | 2.32±0.65 | 2.43±0.55 | **1.31±0.34** |

Table 3 presents the average RMSE over 73 experimental sessions. This result further revealed that the proposed ensemble significantly outperformed the single model, especially for the data of Cluster 2 (single: 2.23±0.93 vs. ensemble with fixed weights: 1.88±0.72 vs. ensemble with dynamic weights: 1.36±0.68) and Cluster 3 (single: 2.37±0.81 vs. ensemble with fixed weights: 1.79±0.62 vs. ensemble with dynamic weights: 1.56±0.52) (sign-rank test, FDR-corrected p-value<0.05).

Fig. 8 presents the prediction results of one test session (that is the 17th session): the red trace is the recorded RT; the purple trace is $\widehat{RT}_{sgl}$, predicted using a single model (single SVR); the yellow trace is $\widehat{RT}_W$, predicted using an ensemble with a fixed weight, which is the averaged posterior probability of the theta power of the first five-minute window in GMM; the blue trace is $\widehat{RT}_{dynW}$, predicted using an ensemble with a dynamically adaptive weight, which is updated every second (as presented in Fig. 8A). The results in Fig. 8A reveals that Model1 had a high weight when the recorded RT is small (such as the time points around 60s, 900s, 2000s, and 3100s). As the recorded RT increased (such as the time points around 250s and 2500s), the weight shifted toward Model3. Fig. 8B compares the recorded RT with the predicted RT. The RTs predicted by the single model ($\widehat{RT}_{sgl}$) were flat (RMSE=3.95), suggesting that this non-adaptive model fails to predict changes in RT. The ensemble with a fixed weight achieved a better prediction performance $\widehat{RT}_W$ (RMSE=1.44) than the single model. However, errors occurred when the driver's RT was high, suggesting that Model1 was not appropriate under the high RT condition and its weight should be changed. The proposed ensemble yields the best prediction result $\widehat{RT}_{dynW}$ (RMSE=1.07). The above demonstrates the ability of the dynamically weighted ensemble to model adaptively a driver's RT.

*E. Reaction Time predicted using Different Models*

This section further shows the extent to which a system could improve the accuracy of predicting RTs if assigning a test sample to a correct cluster of EEG-RT relationship, and vice versa.

As shown in Table 4, when the model was trained using the data of Group 1 (i.e., Model 1) and applied to the EEG data of Group 1, Group 2, and Group 3, the RMSEs obtained were 0.93±0.43, 1.74±0.49, and 1.71±0.45, respectively. The RMSEs obtained when the model was trained using the data of Group 2 (i.e., Model 2) and applied to the EEG data of Group 1, Group 2, and Group 3, were 2.26±0.75, 1.33±0.32, and 2.34±0.46, respectively. Finally, when the model was trained using the data of Group 3 (i.e., Model 3) and applied to Group 1, Group 2, and Group 3, the RMSEs observed were 2.32±0.65, 2.43±0.55, and 1.31±0.34, respectively.

## V. DISCUSSION AND CONCLUSIONS

*A. Key findings*

The trials of lane drift occur at all levels of drowsiness and inattention. In our previous study [55], we only used the single SVR model to predict the RT, which did not consider the changes in fatigue levels during the driving experiment. However, in this study, we applied a GMM model to cluster the fatigue levels of each participant by EEG powers, and then develop a dynamically weighted ensemble-based SVR model to predict the RT.

An ensemble approach [56] for tracking non-stationarity in EEG signals was recently developed as part of a brain-computer interface. It comprised an ensemble of sub-systems, each of which was trained using data recorded from other users. Therefore, a reasonable output could be obtained by fusing the outputs from a set of diverse systems and significant improvement in the overall system performance could be realized when non-stationarity occurred. Many weighted ensemble methods have been proposed; their predictions are based on a weighted vote, with the weights proportional to the resubstituting accuracies. Although the effectiveness of this dynamically weighted ensemble was demonstrated through an empirical comparison with other state-of-the-art models, such an adaptive algorithm typically requires knowledge of the ground truth, making real-world applications of the system unfeasible.

This work proposes an ensemble-based approach to design a system for adaptively modeling driver RTs. The primary purpose is to establish a set of different sub-systems each using the data from a group with similar EEG-RT relationships. The weights determined by posterior probabilities obtained using the Gaussian mixture are used to obtain results for the ensemble members. For the prediction of RTs, the proposed model is found to be significantly better than a single model without an adaptive design (Table 3).

The group labels are initialized in accordance with a subject's overall task performance and re-labeled using a recursive updating procedure. This method enables the EEG-RT relationship to be automatically clustered into various groups, directly contributing to the success of the proposed system. Therefore, the best performance is achieved using the model which is trained using the data with similar EEG-RT relationship. Another key component of the method is the effective model weight assignment. Interestingly, the theta activity that is recorded in the first five minutes of the experiment is associated with the EEG-RT relationship. Using the theta band to compute model weights yields better results than using any of the other bands. These classification results support the findings of an earlier study that the inter-subject frontal midline theta activity is highly related to the variability in the cognitive states [57].

### B. Prediction of Reaction Time using Ensemble Approach

Various phenomena could induce homeostatic changes in the brain, affecting the observed relationship between brain dynamics and behavioral performance, and consequently degrading the performance of a single prediction model. The proposed ensemble is based on the hypothesis that the performance of the system could be improved if the ensemble approach to RT prediction can account for the variability of the EEG-RT relationship. In Table 4, the RMSEs on the diagonal were significantly smaller than those off the diagonal (paired t-test, FDR-adjusted $p$-value<0.01), supporting the hypothesis that model selection importantly affects the system performance in modeling a driver's RT.

### C. Prediction of Reaction Time using a Suitable Model

From Fig. 8A, we find that $Model_1$ that was trained using data for Group 1 (the group with optimal task performance) had the largest weight when the recorded RT was small (e.g., the time points around 60s, 900s, 2000s, and 3100s). As the recorded RT increased (e.g., the time points around 250s and 2500s), the weight provided by $Model_3$, trained using data for Group 3 (the group with poor task performance), is also increased. $Model_1$ is thus inferred to be favorable when a driver is in a 'non-fatigue' state, and $Model_3$ is favorable when a driver is in a 'fatigue' state. However, owing to a lack of objective measurements, the degree of fatigue cannot be clearly identified. Objective measurements need to be defined in the future for this. For instance, in one of our recently completed research projects, an Actigraphy wristband [58] was used to monitor daily variations in subjects' sleep patterns and perceived levels of fatigue on a daily basis. The behavioral results reveal that RT in response to a simulated lane-change event increased with the degree of fatigue. Additionally, the EEG activity reveals that the brain-behavior relationship is significantly correlated with fatigue. This wristband is expected to be useful in elucidating a driver's cognitive state and in selecting an optimal prediction model. This device may facilitate the tracking of a subject's cognitive state. In addition to the EEG power spectra, features of the coupling between remote brain regions, including coherence and connectivity, can be utilized in assigning model weights to improve its performance.

### D. Real-World Applications

This paper provided a breakthrough in neuroergonomics for enhancing driving safety. The developed algorithm for monitoring human performance can offer a new approach for automobile manufacturers in designing new cars. By integrating a feedback mechanism with the proposed algorithm, a closed-loop performance management system can be realized to continuously detect the brain signatures of cognitive lapses and deliver wake-up warnings to operators experiencing momentary cognitive lapses in variable circumstances. Additionally, the developed algorithm can be widely applied significantly benefitting technological developments in the fields of brain-computer interface (BCI), which faces the challenge of the variability in human beings over time, both within the individual and between individuals. The adaptive and effective algorithm developed in this study can model complexities of the dynamic processes in the brain. It can effectively capture the temporal changes in the cognitive state of human subjects, and decipher how the brain codes our behavioral performance.

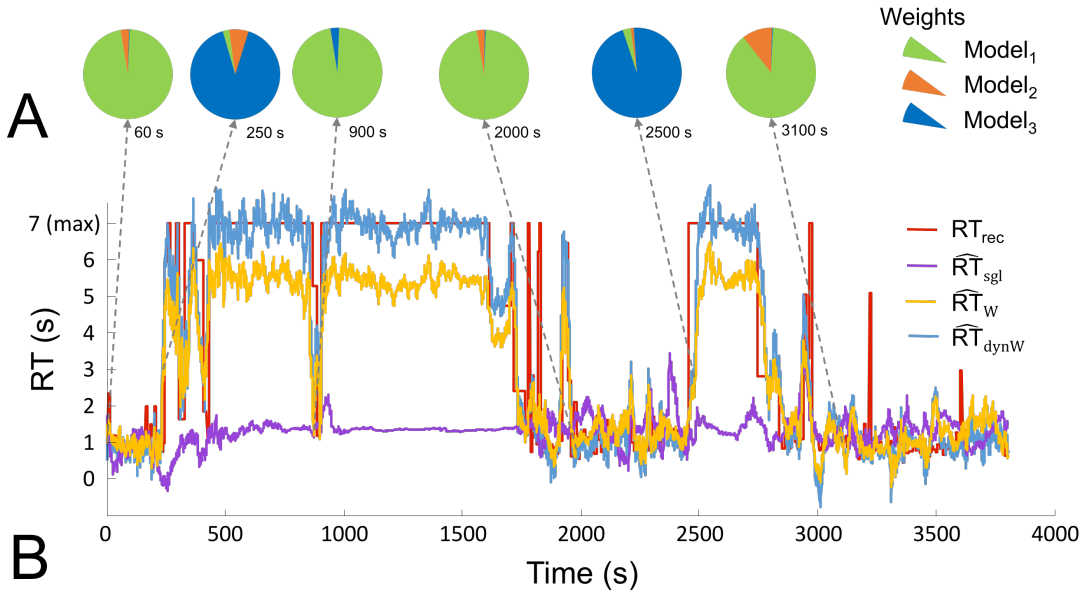

Fig. 8. Prediction results. (A) Dynamical weight changes every second. (B) Comparison between recorded RT (red trace) and predicted RTs (purple trace: $\widehat{RT}_{sgl}$ predicted using a single model; yellow trace: $\widehat{RT}_W$ predicted using proposed ensemble with a fixed weight; yellow trace: $\widehat{RT}_{dynW}$ predicted using proposed ensemble with a dynamic weight).

To examine the feasibility of the proposed system under near real-world circumstances, we have not applied any advanced signal processing method, except for a bandpass filter before FFT, to the data to minimize artifacts. As shown in Table 3, the proposed system is able to reach RMSE in the range between 1.12 and 1.56. The proposed system demonstrated its superiority over conventional single model, but the prediction performance is still far from a satisfactory result.

Several limitations of this model are summarized here. First, we have not used any systematic criterion for deciding on the number of components in the GMM model. A popular approach to address this problem is the use of Bayesian Information Criterion (BIC) [59], which penalizes models with a large number of clusters to avoid over fitting. Since the focus of this investigation is different, we do not use any such criterion, but use an exhaustive grid search for a user-supplied maximum number of clusters. The second issue is that we have used our domain knowledge to decide on the number GMMs. Exploring the structure of the data distribution may enable an alternative way to find a better choice for the number of GMMs to be used. Third, GMM might fail to work satisfactorily if the dimensionality of the features is too high. Some clustering algorithms for variables such as ClustOfVar [59] or structure preserving unsupervised feature selection methods [60] can be used for dimension reduction purpose before mixture modeling. Furthermore, our proposed approach is tested only in an experimental scenario with relatively controlled conditions. As physiological responses may vary wildly between driving situations [60], we cannot guarantee that it would truly exhibit impressive performance in a more uncontrolled scenario. To further improve the prediction performance, application of a real-time artifact removal method to EEG signal and an adaptive model for driving adjustment could be effective.

## ACKNOWLEDGMENTS

This work was supported in part by the Australian Research Council (ARC) under discovery grant DP150101645, in part by Central for Artificial Intelligence, UTS, Australia, and in part by the Army Research Laboratory under Cooperative Agreement W911NF-10-2-0022 and W911NF-10-D-0002/TO 0023. The views and the conclusions contained in this document are those of the authors and should not be interpreted as representing the official policies, either expressed or implied, of the Army Research Laboratory or the U.S Government. The U.S Government is authorized to reproduce and distribute